\RequirePackage[l2tabu, orthodox]{nag}  

\documentclass[10pt]{article}
\usepackage{setspace}
\usepackage{amsmath,amsthm,amssymb,amsfonts} 
\usepackage[pdftex]{graphicx}
\usepackage{float}
\usepackage{subfig}
\usepackage{siunitx}  
\usepackage{booktabs} 
\usepackage[table,xcdraw]{xcolor}
\usepackage{multirow} 
\usepackage{color}
\usepackage{algorithm}
\usepackage{algorithmic} 
\usepackage{microtype} 
\usepackage[square,sort,comma,numbers]{natbib}
\usepackage{lipsum}  
\usepackage{bm}
\usepackage{multicol, blindtext}
\usepackage{authblk}
\usepackage{blindtext}
\usepackage{hyperref}
\usepackage[noabbrev]{cleveref}

\topmargin by 4 \advance \topmargin -1.0in \leftmargin 216mm
\advance \leftmargin -\textwidth \divide \leftmargin by 2 \advance
\leftmargin

\pdfoutput=1

\twocolumn

\newcommand{\etal}{\emph{et al.}}

\def\keywords#1{\gdef\@keywords{\hsize20pc%
		\parindent 0pt\noindent\ignorespaces%
		{ \section*{Keywords} } \ignorespaces #1}
}

\title{{\bf A Survey on Non-Intrusive Load Monitoring Methodies and Techniques for Energy Disaggregation Problem}}
\author[1]{Anthony Faustine \footnote{sambaiga@gmail.com}}
\author[2]{Nerey Henry Mvungi \footnote{nhmvungi@udsm.ac.tz}}
\author[1]{Shubi Kaijage\footnote{shubi.kaijage@nmaist.ac.tz}}
\author[1]{Kisangiri Michael \footnote{kisangiri.michael@nmaist.ac.tz}}
\affil[1]{Dept. of Communication Science and Engineering, NM-AIST, Tanzania}
\affil[2]{College of Information and Communication Technologies, University of Dar es Salaam}

\begin{document}
	
	\maketitle
	

		\begin{abstract}
		The rapid urbanization of developing countries coupled with explosion in construction of high rising buildings and the high power usage in them calls for conservation and efficient energy program.~Such a programme require monitoring of end-use appliances energy consumption in real-time.
		
		The worldwide recent adoption of smart-meter in smart-grid, has led to the rise of Non-Intrusive Load Monitoring (NILM); which enables estimation of appliance-specific power consumption from building's aggregate power consumption reading.~NILM provides households with cost-effective real-time monitoring of end-use appliances to help them understand their consumption pattern and become part and parcel of energy conservation strategy.
		
		This paper presents an up to date overview of NILM system and its associated methods and techniques for energy disaggregation problem. This is followed by the review of the state-of-the art NILM algorithms.~Furthermore, we review several  performance metrics used by NILM researcher to evaluate NILM algorithms and discuss existing benchmarking framework for direct comparison of the state of the art NILM algorithms.~Finally, the paper discuss potential NILM use-cases, presents an overview of the public available dataset and highlight challenges and future research directions.
	\end{abstract}\\
	
	\section{Introduction}
	Residential and commercial buildings consume approximately 60\% of the world’s electricity \footnote{The United Nation’s Environment Programme’s Sustainable Building and Climate Initiative (UNEP-SBCI)}.
	~In U.S. A for example 74.9\% of the produced electricity is used just to operate buildings \footnote{United States Energy Information Administration report:http://www.eia.gov/todayinenergy/detail.cfm?id=14011} while it is 56\% in Africa \citep{Kitio2013}.
	~It is estimated that 80\% more buildings will be in place by 2050\footnote{The United Nation’s Environment Programme’s Sustainable Building and Climate Initiative (UNEP-SBCI)}.
	~Hence energy saving in buildings will have significant impact on the reduction of overall energy demand.
	
	Effective and efficient energy saving in residential buildings can be achieved through real-time monitoring of end-use appliances consumption and provision of real-time actionable feedback to households that give insight into what appliances and when they are used, how much power they consume and why such consumption.~Hence, households will be actively engaged and determine where energy is wasted and where or how to apply the most effective energy saving measures stimulating energy saving behaviour.

	~Studies report that energy consumption awareness coupled with real-time actionable feedback to households inspire positive behavioural change and engage households toward sustainable energy consumption \citep{Batra2015b,Batra2016}.

	Traditionally, real-time appliance-specific breakdown of energy consumption is obtained by deploying sensors (smart plugs) that monitor the consumption of each appliance in buildings.~Deploying such a sensing infrastructure is costly, intrusive and require proprietary communication protocols \citep{Beckel2014,Zhong2013}.~Recently, large scale deployments of smart meters have rekindled the interest towards developing effective non-intrusive load monitoring (NILM)\footnote{Sometimes referred to as Non Intrusive Load Appliance Monitoring (NILAM) or energy disaggregation} techniques \citep{ReyesLua2015}. 
	
	NILM is the computational techniques that use aggregate power data monitored from single point source such as smart meter to infer the end-appliances running in the building and estimate their respective power consumption.
	~It provides households with cost-effective real-time monitoring of end-use appliances that facilitate  energy conservation actions.
	~Also, NILM system can help policy makers evaluate the effectiveness of their energy-efficiency policies while utility can better forecast demand and enable manufacturers to optimize product design to meet customer need \citep{Froehlich2011}.
	
	The initial NILM approach to residential energy disaggregation was proposed by Hart in the 1990s \cite{Hart1992}. Recently many researchers have published several approaches on energy disaggregation that improve the initial design \cite{Zoha2012, Barsim2014}. Despite several efforts done by previous NILM researchers, there are still several challenges which need to be addressed.~This work presents an up to date overview of NILM system and its associated methods and techniques for energy disaggregation problem.
	
	\section{Energy Disaggregation Problem} \label{sec:nilm}
    Energy disaggregation is a technique that estimates the energy consumed by every individual appliance in a house from a single energy measurement device like a smart-meter.
    ~This technique is gaining popularity due to large-scale smart meter deployments worldwide \citep{ReyesLua2015}.~The advantage of this approach is that it can be used in existing buildings easily without introducing any inconvenience to householders being non-intrusive.
    
    Specifically, the problem of energy disaggregation can be formulated as follows: Given the sequence of aggregate power consumption $\bm{X} = \{X_1,X_2...,X_T \}$ from $N$ active appliances at the entry point of the meter at $\bm{t}=\{1, 2..., T\}$, the task of the NILM algorithm is to infer the power contribution  $y_t^i$  of appliance $i \in \{1,2...N\}$ at time $t$, such that at any point in time $t$, 
    \begin{equation}
    X_t = \sum_{i=1}^N y_t^i + \sigma(t)
    \end{equation} where $\sigma(t)$ represents any contribution from appliances not accounted for and measurement noise.
    ~The key challenge to energy disaggregation problem is how to design efficient generalized NILM algorithm across several buildings that
    can run in real-time using smart-meter.~A typical NILM algorithm consists of the following steps: power signal acquisition, event detection, feature extraction and learning\& inference.
   
   \subsection{Power Signal Acquisition}
   This is the first step for any NILM algorithm and it involves acquiring aggregated load measurement at an adequate rate so that distinctive load patterns can be identified.~Several power meters such as Yomo \citep{Klemenjak2016} and c-meter \citep{Makonin2013d} have been designed to measure the aggregated load of the building.~A cost efficient approach for acquiring aggregate power data is to use smart meters which are currently being deployed as the requirement of smart-grid.
   
   The aggregate power signal from these meters can be recorded at different sampling rate.~The sampling frequency is determined by the measurements and electrical characteristics used by NILM algorithm \cite{Zoha2012}.~The sampling frequency can either be high-frequency or low-frequency.
   
   High-frequency is when sampling rate is in a range of \SIrange{10}{100}{\mega\hertz} for the quantity whose electrical characteristics is to be determined.~Power meters for this range are often custom-built and expensive due to sophisticated hardware \citep{Zoha2012}.~Smart meters belong in the low sampling rate of the power signal which is less than \SI{1}{\hertz}. 
   
  \subsection{Event Detection}
  
  The NILM algorithm needs to detects the appliance operations status (e.g ON and OFF) from the power measurements.
  ~The changes in power levels (like ON/OFF) is done by the detection module.
  ~It is a complex process because of different types of appliance in buildings and the different status to be detected like simple ON/OFF, finite state, constantly ON, and continuously variable status as identified in \cite{Hart1992}.~Based on different event-detection strategies, the current NILM approaches can be classified as event-based or state-based.
  
  \paragraph{Event-Based Approaches:}
  The event-based approaches focus on the state transition edges generated by appliances and use change detection algorithm to identify start and end of an event \cite{Barsim2014,Wong2013}.~The task of change detection algorithm is to detect changes in time-series aggregate load data due to one or more appliance being switched ON/OFF or changing its state.~A review on event detection algorithms used in the NILM literature is presented in \cite{Moura2012}.
  
\begin{figure}[ht]
  	\centering
  	\includegraphics[width=0.5\textwidth]{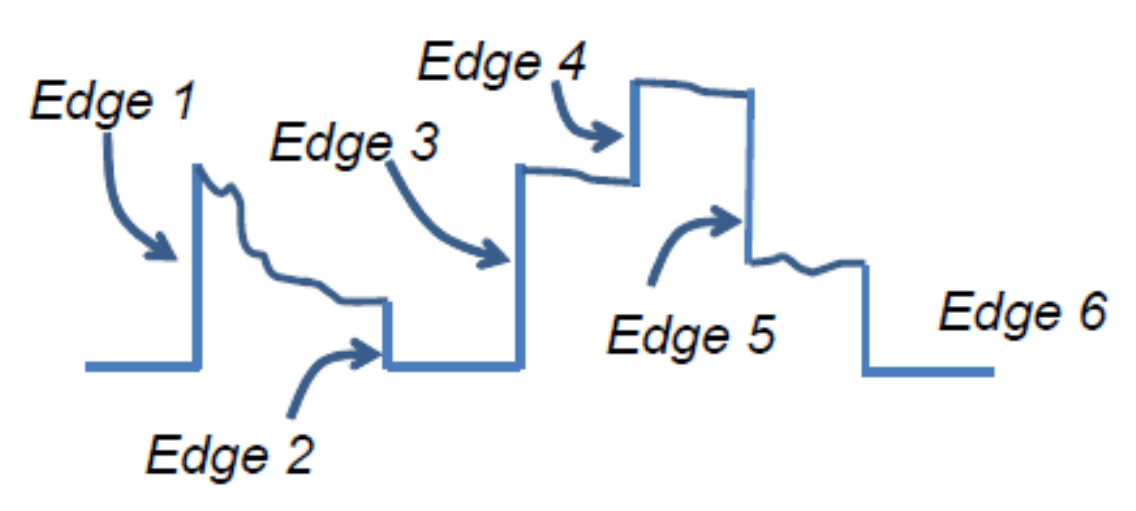}
  	\caption[Edge-based schematic ]{Schematic of Edge-based \citep{ZhaoyiKang2015}}
  	\label{fig:edge}
  \end{figure}

  Event-based approaches rely on the fact that power monitored in  a home is constantly changing (rising and falling, steps) as shown in \autoref{fig:edge}.~These steps (if significant enough) can be an indication that an event has occurred.~Then, appliance signatures such as active power, increasing/falling edge etc are extracted. The extracted appliances signature are analysed to classify the event based on  appliance and its power consumption estimated.~Different classification methods such as Support Vector Method (SVM), neural networks, fuzzy logic, Naive Bayes, k-Nearest Neighbors (kNN), Hidden Markov Model (HMM), decision trees and many other hybrid approaches have been used \citep{Zhao2016a,Altrabalsi}.
  
  The performance of event-based approach is limited by the fixed or adaptive threshold of the change detection algorithm, the large measurement noise, and similarities among steady-state signatures.~In addition, miss detection and false detection of edges may arise in event detection methods. 
  
  \paragraph{State-based Approaches:} 
   State-based NILM-approaches do not rely on event detectors, instead they represent each appliance operation using a state machine with distinct state transition based on appliance usage pattern \citep{Kanghang2016}.~They are based on the fact that when appliance turns ON/OFF or changes running states, create different edge measurements which have a probability distribution that  match to that appliance.~State-based NILMs are usually based in HMM and its variants \citep{Kim2011,Kolter2012,Parson2012,Makonin2015} .

   State-based approaches are limited by the need for expert knowledge to set a-prior value for each appliance state via long periods of training.~Besides, they have high computational complexity \citep{Kanghang2016,Zoha2012} and do not have a good way to handle the fact that states may stay unchanged for long time intervals \cite{Mauch2016}.

  \subsection{Feature Selection}
  Effective NILM algorithm requires unique features or signatures that characterize appliance behaviour.~All appliances type have a unique energy consumption pattern often termed as “appliance signatures”.~This unique energy consumption pattern is often used to uniquely identify and recognize appliance operations from the aggregated load measurements \citep{Zoha2012}.~According to \cite{Hart1992}, appliances features are measured parameter of total load that give information about nature or operating state of an individual appliance in the load.~It is unique consumption pattern intrinsic to each individual electrical appliance \cite{Liang2010}.~There are two main classes of appliance signatures used by NILM research for appliances identification namely transient features and steady-state features.
  
  \paragraph{Transient signatures:} Are short-term fluctuations in power or current before settling into a steady-state value. These features uniquely define appliance state transitions by extracting features like shape, size, duration and harmonics of the transient \citep{Zoha2012}.~They require high sampling rates to obtain a high degree of signal uniqueness and longer monitoring time in order to capture all operation cycles \cite{Kanghang2016}.~This in turn, demands a costly hardware to be installed in households since smart meters reports only low-frequency power.~For example, Patel \etal~\cite{Patel2007} use a custom built hardware to detect the transient noise from \SIrange{0.01}{100}{\kilo\hertz}.~The authors use the fact that each appliances in state operation transmits noise back to the power line.
  
  \paragraph{Steady state features:} Relate to more sustained changes in power characteristics when an appliance  change its running states.~These features include; active power \citep{Liao2015,Parson2012}, reactive power \citep{Zeifman2011b}, current \citep{Makonin2013}, current and voltage
  waveforms \citep{Barsim2016}; just to mention a few.~The extraction of steady-state signature does not demand high-end metering devices and can be obtained from RMS values of current and voltage.~Steady-state features are the most commonly used features at low frequency in the literature.~While most of prior works such as \cite{Parson2012,Kolter2012,Zhao2015} use real-power for disaggregation, \cite{Makonin2013} argue that  current, rather than real power, is a more effective steady-state feature for energy disaggregation problem.

  \subsection{Learning and Inference in NILM}
  
  In this stage the extracted appliances signature are analysed in order to classify an appliance specific states and estimate its corresponding power consumption.~The learning algorithms are used to learn the model parameters while the inference algorithms are employed to infer appliance states from observed aggregate power data and estimate their corresponding power consumption.~The algorithm for learning in NILM can be supervised or unsupervised.
  
  Supervised NILM techniques require a training phase in which both the aggregate data and the individual appliance consumption are used.~In that case, sub-metered appliances data or labeled observations must be collected from the target building.~The process of collecting these data is expensive, time-consuming and limit the scalability of NILM systems \citep{Barsim2015}.~Several existing works have focused on supervised learning techniques such as Support Vector Machine(SVM)\citep{Altrabalsi2016,Presented2014}, Nearest Neighbour(k-NN) \citep{Altrabalsi} and some forms of HMM \citep{Li2014}.
  
  Unlike supervised NILM, unsupervised NILM techniques do not require pre-training and thus suitable for real time NILM application. Unsupervised NILM approaches do not require individual appliance data, the models parameters are captured only using the aggregated load, without the user intervention \citep{Bonfigli2015}. Current NILM research has focused on building unsupervised learning models which are less costly and more reliable \citep{Bonfigli2015,Zoha2012}. 
  
  Unsupervised NILM approaches can further be grouped into three subgroups as suggested by \cite{Zhao2016a}; 
  First are the unsupervised approaches that require unlabelled training data to build appliance model or populate appliances database.~They are usually based on HMM and the appliance models are either generated manually \citep{Makonin2015} or automatically \citep{Parson2012} during the training phase.~Most of these approaches can not be generalized into unseen buildings.
  
  The Second groups includes unsupervised approaches that use labelled data from known house to build appliances models which are then used for disaggregation in unknown (unseen) building. These approaches require sub-metered appliances data to be collected from the training or known house.~These data are used to build generic appliance models which is then used in unseen buildings.~Most deep learning based NILM techniques such as in \cite{Kelly2016a} fall in this category. 
  
  Lastly are the unsupervised approaches that do not require training before energy disaggregation takes place.~These approaches can perform energy disaggregation without the need of sub-metered data or the prior knowledge \citep{Zhao2016a,Jia2015}.

	\section{State-of-the-arts NILM Algorithms}\label{sec:algm}
	Several state-of-the-art NILM unsupervised algorithms have been proposed using different approaches such as different variants of HMM \cite{Kim2011,Parson2012,Kolter2012,Makonin2015}, Graph Signal Processing (GSP) \cite{Stankovic2014,Zhao2016a} and Deep earning \cite{Badayos2015,Paulo2016a}. 
	
	\subsection{Hidden Markov Model}
	HMM is a Markov model whose states are not directly observed instead each state is characterised by a probability distribution function modelling the observation corresponding to that state \citep{Kim2011,Rabiner1989a}.~There are two variables in HMM: observed variables and hidden variables where the sequences of hidden variables form a Markov process.~In the context of NILM, the hidden variables are used to model appliances states (ON,OFF, standby etc) of individual appliances and the observed variables are used to model the electric usage.~HMMs has been widely used in most of the recently proposed NILM approach because it represents well the individual appliance internal states which are not directly observed in the targeted energy consumption.
	
	A typical HMM is characterised by the following: The finite set of hidden states $S$ (e.g ON, stand-by, OFF, etc.) of an appliance, $S = \{S_1, S_2....,S_N\}$. The finite set of observable symbol $Y$ per states (power consumption) observed in each state, $Y = \{y_1, y_2....,y_T\}$. The observable symbol $Y$ can be discrete or a continuous set.~The transition matrix \textbf{A} $=\{a_{ij},1\leq i,j \geq N\} $ represents the probability of moving from state $S_i$ to $S_j$ such that:
	$a_{ij} = P(q_{t+1} =S_j \mid q_t=S_i)$, with $a_{ij} \leq 0$ and where $q_t$ denotes the state occupied by the system at time $t$.~The emission matrix \textbf{B}$ =P(y_t \mid S_j) $ representing the probability of emission of symbol $y_t$ $\epsilon$ $Y$ when system state is $S_j$.~The initial state probability distribution is $\bm{\pi} = \{\pi_i \}$ indicating the probability of each state of the hidden variable  at $t = 1$ such that, $\pi _i = P(q_1 = s_i), 1 \leq i \geq N$.~The set of all HMM model parameters is represented by 
	$\bm{\lambda =\{\pi, A, B \}}$.
	
	When applying HMM to a real world problem, two important problem must be solved.~First how to learn the model parameter $\bm{\lambda}$ given the sequences of observable variable $\bm{Y}$.~Second, given the model parameter $\bm{\lambda}$ and the sequences of observable variable $\bm{Y}$ how to infer the optimal sequences of hidden state $\bm{S}$.~These problems are referred to as learning and inference  problems.~Various algorithms such as Baum-Welch algorithm and the Viterbi algorithm  have been proposed to solve these problems.
	
\begin{figure}[ht]
		\centering
		\includegraphics[width=0.4\textwidth]{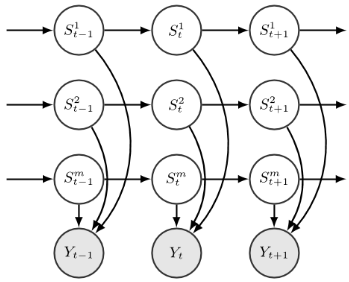}
		\caption[Graphical representation of FHMM]{Graphical representation of FHMM}
		\label{fig:FHMM}
	\end{figure}

	The factorial HMM (FHMM) is an extension of HMM with multiple independent hidden state sequences and each observation is dependent upon multiple hidden variables \citep{Z.Ghahramani1995}. In FHMM, if we consider     
	$\bm{Y} = \{y_1,y_2,.....y_T\}$ to be the observable sequences then $\bm{S = \{S^{(1)},S^{(1)},.....S^{(M)}\}}$ represents the  set of hidden state sequences where $\bm{S^{(i)}} = \{S^{(i)}_1,S^{(i)}_2,.....S^{(i)}_T\}$ is the hidden state sequence of the chain $i$ as shown in  \autoref{fig:FHMM}.
	FHMM are preferred to HMMs for modelling time series generated by the interaction of several independent process. However, the computational complexity of both learning and inference is greater for FHMMs compared to HMMs. In addition, the inference techniques for FHMM based approach is highly susceptible to local optima \citep{Zoha2012}. 
	
	Several HMMs based NILM algorithms for energy disaggregation at low sampling rate has been proposed in the literature.~In \cite{Kim2011} unsupervised technique for energy disaggregation using a combination of four FHMM variants is proposed.~The authors use low-frequency real power feature and assume a binary state of appliances (ON and OFF state only).~To learn model parameters, Kim's approach uses Expectation Maximasation (EM) algorithm  and employ Maximum Likelihood Estimation(MLE) algorithm to infer load states.~The performance of Kim's technique is limited to few number of appliances, require appliances to be manually labelled after disaggregation and suffer from high computational complexity which makes it unsuitable for real-time applications \citep{Parson2014c}. 
	
	The work presented in \cite{Kolter2012} propose a new inference algorithm for unsupervised energy disaggregation called Additive Factorial  Approximate MAP~(AFMAP) that is computationally efficient and does not suffer from local optima.~The AFMAP algorithm is used to perform approximate inference over the additive FHMM.~However, the model requires  appliances  to be manually labelled after off-line disaggregation  and have a low performance for electronics and kitchen appliances.

	Parson \etal~\cite{Parson2012}, introduce an approach that use difference HMM from \cite{Kolter2012} as Bayesian network for disaggregation of active power with \SI{60}{\second} sampling rate.~To perform inference, the authors use an extension of viterbi algorithm and propose an EM training process to build a generic appliance model for learning the model parameters.~This generic model is then tuned to specific appliance instances using only aggregate data from home in which NILM is being applied.~The active tuning process requires a training window of data where no other appliance changes state.~For the cyclic types of appliance such as fridges, this is easy since it is often the only appliance running at night, but it is generally difficult for other appliances \cite{Huss2015}.
	
	A fully unsupervised NILM framework based on non-parametric FHMM using low-frequency real power feature is presented in \cite{Jia2015}. They use the combination of slice sampling and Gibbs sampling to do inference that simultaneously detect number of appliances and disaggregate the power signal from the composite signal.~However, for larger disaggregation problems this inference algorithm becomes a limitation as it may stuck in local optima \citep{Gael2011}.~Besides it difficult to see this algorithm runs in real-time owing to complexity problem of FHMM.
	
	Makonin \etal~\cite{Makonin2015} present another NILM algorithm for low-frequency sampling rate that uses a super-state HMM in which a combination of modeled appliances states is  represented as one super state.~The authors propose a new variant of viterbi algorithm called sparse Viterbi algorithm.~This algorithm  perform computationally efficient exact inference instead of relying on approximate inference method like in FHMM based approach.~Makonin's approach preserves dependencies between appliances, can disaggregate appliances with complex multi-state power signatures and can run in real-time on an inexpensive embedded processor.~Although the reported approach can disaggregate large number of super-state, there is  still a limitation in time and space since number of super-states grow exponentially with the number of appliances. 
	
	Despite the fact that HMM-based NILM approaches have been widely used in energy disaggregation they require an expert knowledge to set  a-priori values for each appliance state.~Their performance are thus limited by how well the generated models approximate appliance true usage \citep{Zhao2016a}.~Moreover, HMM-based approaches have better performance for controlled multi-state appliances like refrigerator, but their performance degrades for uncontrolled multi-state and variable appliances \citep{Mauch2016}.
	
	\subsection{Graph Signal Processing}
	
	Graph Signal Processing (GSP) or signal processing on graph is an emerging field that extends classical signal processing theory to data indexed by general graphs \citep{Sandryhaila2014}.~GSP represents a dataset using a graph signal defined by a set of nodes and a weighted adjacency matrix \citep{Zhao2015}.~Each node in the graph corresponds to an element in the dataset while the adjacency matrix  define all directed edges in the graph and their weights, where assigned weights reflects degree of similarity or correlation between the nodes \citep{Stankovic2014}.~GSP is the powerful, scalable and flexible signal processing approach that is suitable for machine learning and data mining problems.~In particular, GSP is suitable for data classification problems in which training periods are short and inefficient to build appropriate class models \citep{Sandryhaila2013b}.
	
	Given a set of aggregate power measurement $X$ we define  a graph $G = \{\mathcal{V}, A  \}$ where $\mathcal{V}$ is the set of nodes corresponding to the acquired measurements and $A$ is the weighted adjacency matrix of a graph which define the edge of a graph.~Each element $x_i \in X$ corresponds to a node $\mathit{v_i}\in\mathcal{V}$ and each weight $A_{ij}$ of the edge between nodes $\mathit{v_i}$ and $\mathit{v_j}$  reflects the degree of relation  between $x_i$ and $x_j$.~The weight of a node $A_{ij}$ is  usually defined using gaussian kernel weighting function, the most used kernels in machine learning for expressing similarities between dataset defined by \cref{eq:kernel}.
	\begin{equation}\label{eq:kernel}
	A_{ij} = \exp \Big[ -\frac{(x_i-x_j)^2}{\rho^2}\Big] 
	\end{equation}
	where $\rho$ is a scaling factor \citep{Kanghang2016}.~A graph signal is then defined as a map from  the map on the graph nodes $\mathcal{V}$ to the set of complex number $\mathbf{s}$ where each element $s_i \in \mathbf{s}$ is indexed by nodes $v_i \in\mathcal{V}$ \citep{Sandryhaila2014}. In the context of energy disaggregation, each vertex $v_i \in \mathcal{V}$ is associated to aggregate power variation signal between adjacent power reading one sample $\Delta X_t , $ where $\Delta X_t=X_{t+1} -X_t$.~For further literature on GSP, interested reader may refer to \cite{Sandryhaila2014}.
	
	Recently, researchers have proposed different GSP-based approach for NILM.~The first GSP-NILM approach that is neither state-based nor event-based was presented in \cite{Stankovic2014}.~The authors, leverage on the work by \cite{Sandryhaila2013b} to perform low-complexity multi-class classification of the acquired active power readings without the need for event detection to detect appliance changing states.~However, this approach is supervised and employs GSP only for data classification \citep{Zhao2015}. 
	
	Zhao \etal~\cite{Zhao2015,Zhao2016a} propose a blind, low-rate and steady state event-based GSP approach that does not require any training. The proposed GSP-NILM disaggregate any aggregate active power dataset without any prior knowledge and relay upon the GSP to perform adaptive threshold, signal clustering and pattern matching \citep{Zhao2015,Zhao2016a}.~Zhao's approach work well if the average load of each appliance is distinct enough from other appliances load and if the power of each load does not fluctuate much.~This is not a typical case in most building and thus limit the performance of Zhao's algorithm.~Additionally, the proposed GSP approach require appliances to be manually labelled after disaggregation, highly affected by noise and it's performance is also  limited by the event detection performed via adaptive thresholding \citep{Sandryhaila2013b}.

	\subsection{Deep Learning}
	
	Deep learning is the machine learning approach that has drawn heavily on the knowledge of the human brain (artificial neural networks), statistics and applied mathematics \citep{Bengio2016Book}.~It is the artificial neural networks (ANN) that are composed of many layers.~For a comprehensive survey and more details on deep-learning interested reader should refer to \cite{Deng2014,Bengio2016Book}.
	
	In recent years, deep learning has made substantial improvements in several fields such as computer vision \cite{Taigman2014}, speech recognition \cite{Amodei2015} and machine translation \cite{Yonghui2016}.~This is mainly due to more powerful computers, larger datasets and techniques to train deeper networks.~In addition, deep learning models are flexible (enabling similar models to be used in wide range of  problems) \cite{Deng2014}.~Recently, different deep learning architecture such as Recurrent Neural Network (RNN) \cite{Badayos2015}, Convolutional Neural Network (CNN) \cite{Badayos2015,Zhang2016,Paulo2016a}, Auto encoder \cite{Badayos2015} and a combination of deep learning and HMM \cite{Zhang2016,Mauch2016b,Huss2015} has been employed to the energy disaggregation problem.
	
	A deep learning novel approach for energy disaggregation that identifies additive sub-components of the power signal in an unsupervised way is presented in \cite{Lange2016}.~The approach uses high-frequency measurements of current, assume two-state appliances models and requires buffering of all the data until inference.~Barsim \etal~\cite{Barsim2016} propose neural network ensembles approach to address NILM problem. The ensemble of neural networks are used in appliance identification problem from the raw high-resolution current and voltage waveforms.~The work by Kelly \etal~\cite{Kelly2016a}, adapted three neural network architectures to low-frequency energy disaggregation problem.~In \cite{Paulo2016a}, Paulo \etal~present a comparison of variety of CNN and RNN for energy disaggregation across a number of appliances.
	
	The work presented by \cite{Huss2015} uses CNN network to extract appliance features which are then used as observations to a  hidden semi Markov model (HSMM).~Huss's model performed considerably better than a CNN alone with reduced computational cost.~In \cite{Mauch2016}, Mauch \etal~propose a novel combination of HMM with deep neural network (DNN) for load disaggregation.~Mauch's approaches trains HMM with two emission probabilities, one for single load to be extracted which is modelled as a Gaussian distribution and other for the aggregate signal in which DNN is used.~Despite the fact that Mauch's DNN-HMM models outperformed FHMM, it was trained using few data (20.7 days REDD data).~Deep learning models require lots of data in order to be well generalized.
	
	Zhang \etal~\cite{Zhang2016} propose a sequence-to-point learning with CNN for energy disaggregation in which a single-midpoint of an appliance window is treated as classification outputs of a neural network with the mains window being the input.~This differ from the work presented in \cite{Badayos2015} in which a given window of the main sequence is treated as input and a sequence of target appliance as the output of the neural network.~The authors further integrate CNN and AFHMM using logarithmic opinion pool method.~Zhang approach was found to outperform HMM based approaches \cite{Kolter2012,Zhong2015} and  deep learning approach by Kelly \etal~\cite{Badayos2015} with reduced computational cost.
	
	\section{Evaluating NILM Algorithms}\label{sec:metrics}
	
	Defining relevant evaluation standards such as performance metrics and benchmarking framework are crucial to enable empirically evaluation of NILM algorithms and get fair performance comparison between algorithms. 
	
	\subsection{Performance Metrics}
	
	NILM researchers use several performance metrics to evaluate  energy disaggregation algorithms.~To measure how well an algorithm can  predict how an appliance is running in each state(switching ON or OFF), several NILM researchers use accuracy metric defined in \eqref{eq:acc}.
	
	\begin{equation}\label{eq:acc}
	Acc. = \frac{correct\_matches}{total\_possible\_matches}
	\end{equation}
	Since appliance usages in a house is a relative rare event, accuracy metric is sometimes misleading. Accuracy metric is not descriptive for an appliance that is mainly OFF \citep{Huss2015,Kim2011}. For example, if a TV is ON 10\% of the time, an algorithm that predict that the TV is always off will have 90\% accuracy without any ability to predict its usage.~As results, classification accuracy measures, such as F-Measure ($F_M$) has been used \citep{Altrabalsi2016,Iwayemi2015,Stankovic2014}. 
	
	F-Measure is the harmonic mean of precision (PR), the positive predictive values and recall (RE) which is the true positive rate or sensitivity defined by \crefrange{eq:fm1}{eq:fm2}
	
	\begin{align}
	PR &= \frac{TP}{(TP + FP)} \label{eq:fm1} \\
	RE &= \frac{TP}{(TP + FN)} \\
	F_M & = \frac{2\times PR \times RE}{(PR + RE)}\label{eq:fm2}
	\end{align} where true positive (TP) presents the correctly detected state of ON or OFF, false positive (FP) represents an incorrect detection that is predicted appliance was ON while OFF, and false negative (FN) indicates that the appliance used was not identified i.e. was ON but not detected. However, F-measure is limited to binary appliances (OFF/ON) and not applicable for multi-state appliances \citep{Makonin2014b}.
	
	To  account for multi-state appliances, Makonin \etal~\cite{Makonin2014b} introduce a finite-state F-Measure (FS-$F_M$) by adapting the work by \cite{Kim2011}.~Their approach split TP into two: inaccurate true-positives (ITP) and accurate true-positives (ATP). The ITP is a partial penalization measure which converts the binary nature of TP into a misclassification that is not binary in nature defined by \crefrange{eq:itp}{eq:atp}
	\begin{align}
	ITP &= \frac{ \sum_t^T\mid\hat{s}_t^i -s_t^i \mid }{K^i} \label{eq:itp}\\
	ATP &= 1 - ITP \label{eq:atp}
	\end{align} where $\hat{s}_t^i$ is the estimated state from appliance $i$ at time $t$, $s_t^i$ is the ground truth state, and $K^i$ is the number of states
	for appliance $i$. To take account for these
	partial penalizations, \cite{Kim2011} redefined precision (PR) and recall (RE) stated  as \crefrange{eq:pr}{eq:re}.
	\begin{align}
	PR & = \frac{ATP}{(ATP + ITP + FP)} \label{eq:pr} \\
	RE & = \frac{ATP}{(ATP + ITP + FN)}  \label{eq:re}
	\end{align}
	The FS-$F_M$ remains the harmonic mean of the new precision and recall. Accuracy and F-Measure are called classification metrics because they only measure how accurately NILM algorithms can predict what appliance is running in each state.
	
	To measure how well NILM algorithm is able to estimate and assign the power consumed by each appliance different measures have been used. Root mean square error (RMSE) is one of such estimation accuracy measure that NILM researchers use \citep{Batra2013,Parson2012,Batra2015b}. The RMSE between the estimated power $\hat{y}_t^i$ consumed at time $t$ for appliance $i$ and  the ground truth power $y_t^i$ consumed at time $t$ for appliance $i$ is given by \cref{eq:rmse}
	\begin{equation}\label{eq:rmse}
	RMSE = \sqrt{\frac{1}{T} \sum_t (\hat{y}_t^i - y_t^i)}
	\end{equation}
	where $T$ denotes the number of samples recorded. Using RMSE measure is hard to compare how the disaggregation of one appliance performed over
	another since this measure is not normalized \citep{Makonin2014c}. To address this issue, other researchers \citep{Kolter2012, Lange2015, Batra2013} use  the normalized disaggregation error that provides a normalized measure of the difference between the actual and the estimated power consumptions of the $i^{th}$ appliance given by equation \cref{eq:de}.
	\begin{equation}\label{eq:de}
	de =\displaystyle \sqrt{\frac{\displaystyle \sum_{t,i} ||\hat{y}_t^i - y_t^i ||^2}{\displaystyle \sum_{t,i} || y_t^i ||^2}}
	\end{equation} 
	
	The estimation accuracy proposed by \cite{Kolter2011} can be used to evaluate the overall performance of the NILM algorithm and is defined by \cref{eq:eacc} 
	\begin{equation}\label{eq:eacc}
	E_{Acc} = \displaystyle \Bigg[ 1 - \frac{\sum_{t=1}^T \sum_{i=1}^N |\hat{y}_t^i - y_t^i |}{2 \sum_{t=1}^T \sum_{i=1}^N | y_t^i | } \Bigg]
	\end{equation} where $T$ is the time sequence or number of disaggregated readings and $N$ is the number of appliances. Using this metric, we can derive the estimation accuracy for each appliance by eliminating the summations over $N$ as in \cref{eq:eaccn}
	\begin{equation}\label{eq:eaccn}
	E_{Acc}^{(i)} = \displaystyle \Bigg[ 1 - \frac{\sum_{t=1}^T |\hat{y}_t^i - y_t^i |}{2 \sum_{t=1}^T  | y_t^i | } \Bigg]
	\end{equation}
	The disaggregation error (de) in \cref{eq:de} and estimation accuracy for each appliance ($E_{Acc}^{(i)}$) in \cref{eq:eaccn} measure how well the estimated power profiles
	match the actual power profiles overtime \citep{Piga2016}. The low values of the disaggregation error or RMSE (high value of the estimation accuracy) imply an accurate disaggregation.
	
	The work by \cite{Piga2016} introduce another estimation metric called the estimated energy fraction index (EEFI) which provides the fraction of energy assigned to the $i^{th}$ appliance. The EEFI is defined by \cref{eq:eefi}
	\begin{equation}\label{eq:eefi}
	EEFI = \displaystyle \frac{\sum_{t=1}^T \hat{y_t^i}}{\sum_{t=1}^T \sum_{i=1}^N \hat{y_t^i}} 
	\end{equation} and it should be compared with the actual energy fraction index (AEFI) which provides the actual fraction of energy consumed by the $i^{th}$ appliance defined by \cref{eq:aefi} 
	\begin{equation}\label{eq:aefi}
	AEFI = \displaystyle \frac{\sum_{t=1}^T y_t^i}{\sum_{t=1}^T \sum_{i=1}^N y_t^i} 
	\end{equation}
	
	\subsection{Benchmarking}
	The lack of efficient benchmarking framework is another major challenge in the field of NILM research. This has greatly attributed by the fact that there is no reference algorithms implementation. Hence, NILM researchers use different metrics, different datasets, and different pre-processing steps \citep{Kelly2016a,Batra2014a}. It is therefore empirically difficult to perform direct comparison or to reproduce results of the state-of the art NILM algorithms. As the result, the newly proposed approaches are rarely use the same benchmarking algorithms and/or most of the comparison studies are sometime misleading or favour the work presented \citep{Huss2015}.
	
	To address the above mentioned challenge, Batra \etal~\cite{Batra2014a} and Kelly \etal~\cite{Kelly:2014:NVN} developed Non-Intrusive Load Monitoring Toolkit (NILMTK)\footnote{http://nilmtk.github.io/}. NILMTK is an open-source NILM toolkit written in Python and designed specifically to enable the comparison of NILM algorithms across diverse data sets. It contains data set parsers, data set analysis statistics, preprocessors for reformatting data sets, benchmark disaggregation algorithms, accuracy metrics and rich metadata support via the NILM Metadata\footnote{https://github.com/nilmtk/nilm\_metadata}\citep{ParsonFHBKSKR15}. 
	
	The NILMTK toolkit provides a complete pipeline from data sets to accuracy metrics, thereby lowering the entry barrier for researchers to implement a new algorithm and compare its performance against the current state of the art \citep{Batra2014a}. Several studies such as \citep{Badayos2015,Batra2014d,Batra2014b,Batra2015b,Batra2016} have used this toolkit to implement and evaluate their NILM algorithms.~The work presented in \cite{Kelly2015} use NILMTK to analyse and validate the UK-DALE dataset. 
	
	The release of NILMTK was followed by the NILM-Eval framework\footnote{https://github.com/beckel/nilm-eval} developed by ETH Zurich distribution system research group\footnote{http://vs.inf.ethz.ch/}. The NILM-Eval is a comprehensive evaluation framework for NILM algorithms written in MATLAB. It is designed to facilitate the design and execution of large experiments that consider several different parameter settings of various NILM algorithms \citep{Beckel2014}. The NILM-Eval framework allows new NILM researcher to replicate experiments performed by others or evaluate an algorithm on a new dataset and fine-tune configurations to improve the performance of an algorithm in a new setting \citep{Kelly2016a}. In their study, \cite{Li2015} implemented the algorithms on MATLAB and evaluated the performance of their approach on the NILM-Eval framework.
	
	Recently, \cite{Pereira2015} propose an approach that aim to help NILM researchers systematically evaluate and benchmark NILM technology across different datasets and performance metrics, using open source technologies and well-established performance metrics and evaluation techniques. However, the tool is limited to event-based approaches.

	 \begin{table*}[ht]
		\centering
		\caption{Publically Available Energy Dataset Comparison}
		\scalebox{0.60}{
			\begin{tabular}{p{2cm}p{3cm}p{3cm}p{1cm}p{3cm}p{5cm}p{5cm}p{5cm}}
				\toprule
				\textbf{Dataset} & \textbf{Location} & \textbf{Duration} & \textbf{No.of houses} & \textbf{Sensors/house} & \textbf{ Resolution} & \textbf{Features} & \textbf{Other Data} \\
				\midrule
				\rowcolor{black!20} REDD \citep{Kolter2011} & USA & 3-19 days & 6 & 24 & 15KHz(Aggr), 0.5Hz and 1Hz (sub) & V and P (Aggr), P (sub) &  \\
				BERDS \citep{Maasoumy2013} & USA & 1 year & 1 & 4 & 20sec & P,Q and S & climate data \\
				\rowcolor{black!20} BLUED \citep{Anderson2012a} & USA & 8 days & 1 & Aggregated & 12KHz(Aggr only) & I, V and State transition label for each appliance. &  \\
				Smart \citep{Barker2012a} & USA & 3 months & 3 & 21-26 circuit meters & 1Hz & P and S (Aggr), P (Sub) & electricity generation data from on-site solar panels and wind turbines, outdoor weather data, temperature and humidity data in indoor rooms \\
				\rowcolor{black!20} DRED \citep{N2015} & Netherlands & 6 months & 3 & 12 appliances & 1Hz & P (Aggr \& Sub) &  indoor temperature, outside temperature, wind speed, pre-cipitation, humidity and occupancy data. \\
				Tracebase \citep{Reinhardt2012} & Germany & N/A & 15 & 158 devices & 1-10sec(Sub only) & P &  \\
				\rowcolor{black!20} AMPDS \citep{Makonin2013} & Canada & 1 year & 1 & 19 & 1min & V, I, F P, Q, S and Pf & water and natural gas, \\
				AMPds2 \citep{Makonin2016} & Canada & 2 & 1 & 21 & 1min & V, I, F P, Q, S , Pf ,real energy, reactive energy, and apparent energy & water and natural gas, weather data and   utility billing data. \\
			  \rowcolor{black!20}	UK-DALE \citep{Kelly2015} & UK & 499 days, 2.5 years(house 1) & 5 & 5-54 devices & 16 kHz(Aggr) and 1/6 Hz(Sub) & P and switch status &  \\
				iAWE \citep{Batra2013} & India & 73 days & 10 & 33 devices & 1sec(Aggr) and 1sec or 6sec (Sub) & V, I, F, P and phase & Water and ambient conditions \\
				\rowcolor{black!20} REFIT \citep{Stankovic2016} & UK & 2years & 20 & 11 & 8sec & P & Gas and environmental data \\
				GREEND \citep{Monacchi2015a} & Austria/ Italy & 1year & 9 & 9 & 1Hz & P &  \\
				\rowcolor{black!20} ECO \citep{Beckel2014} & Switzerland & 8months & 6 &  & 1Hz & P and Q & Occupancy information \\
				IHEPCDS \footnote{http://archive.ics.uci.edu/ml/datasets/Individual+household+electric+power+consumption} & France & 4 years & 1 & 3 & 1min & V, I, P and Q &  \\
			 \rowcolor{black!20}	OCTES \footnote{http://octes.oamk.fi/final/} & Scotland,Iceland \&Finland & 4–13months & 33 & Aggregated & 7sec & P and phase &  \\
				HES & UK & 1month(255 houses), 1year(26houses) & 251 & 13-51 & 2min & P &  \\
			\rowcolor{black!20}	ACS-F1 \citep{Gisler2013} & Switzerland & 2, 1 hoursessions & NA & 100, 10 types & 10sec & P, Q, I, f, V and phase & \\
				\bottomrule 
			\end{tabular}%
		}
		\label{tab:dataset}%
     \raggedright{{\footnotesize  Aggregte (Aggr), Sub-metering (sub), Active Power (P), Reactive Power (Q), Apparent Power (S), Energy (E), Frequency (f), Voltage (V) and Current (I)}}

	\end{table*}

	\section{Non-Intrusive Load Monitoring Use-Cases}\label{sec:usecase}
	
	Despite the fact that NILM techniques promise several useful applications for energy conservation in buildings, its broader applications have not been fully realized.~This is attributed to the fact that most of NILM researchers have focused on accurate disaggregation and not concrete application.~According to \cite{Barker2014} NILM research needs to move past computing appliance-level energy breakdowns with emphasize on designing new and novel applications that lead to sustainable energy saving in buildings.~Converting the energy disaggregation data into actionable feedback will improve energy efficiency in residential buildings and engage consumers in the path toward sustainable energy in buildings.~NILM researchers should put much of their emphasis in designing new and novel applications rather than seeking incremental improvements in algorithms accuracy. 
	
	One of the most important applications of NILM is the provision of real-time actionable  energy feedback or recommendations to households that could lead to sustainable energy saving.~This information could help households identify unnecessary consumption, identify inefficient appliances and suggest optimisations, raise alerts and make consumers more aware of the energy they consume.~For example, using this information households can detect when appliance be switched to a more energy efficient mode \citep{Parson2014c}.~This greatly helps households not only understand their consumption pattern but also become part and parcel of energy conservation \citep{Batra2014e}. Parson \etal~present an application that use NILM algorithms to infer fridge usage in UK by providing households with feedback on energy-money trade-offs of shifting to new energy-efficient fridges \cite{Parson2012}.~Another application of NILM to a large number of households smart meter data is presented in \citep{Parson2014}.~The authors propose an approach by which the energy efficiency of fridge and freezers are estimated from an aggregate load and calculate the time until the energy savings of replacing such appliances have offset the cost of the replacement appliance.
	
	Temporal pattern such as unusual power consumption in NILM data can be used in detection of fault appliances or malfunctioning appliances in residential buildings.~This information can be further used to determine time to retrofit old appliances and detect degraded performance of appliance in buildings.~In such a situation, households can be provided with real-time alert feedback by either suggesting a more energy efficiency replacement for less expensive appliances or a repair for more expensive appliances.~Through field test of NILM algorithm,  \cite{Hart1992} detected a failed appliance by its abnormal low power consumption and faulty refrigerator which was ON almost all of the time. Batra \etal~ develop and demonstrate techniques that use NILM actionable feedback about a refrigerator and HVAC to residential users \cite{Batra2015b}.~Their application provide targeted actionable feedback with specific actions such as repair or fix to users with much more energy due to fridge usage than normal or fridge that are malfunctioning or miss configured. 
	
	NILM system can also be used to support and enhance continues energy audits in buildings that currently require multiple measurements across buildings.~Energy audit is a process by which a building is inspected and analysed to determine how energy is used with aim to identify opportunity for energy conservation \citep{Berges2010c}.~Detail analysis of NILM data can be used to either suggest ways of reducing consumption and cost or confirm the energy saving resulting from conservation measures.~In \cite{Berges2010c} Berges \etal~present an experimental NILM system for supporting residential energy audits.
	
	NILM can further be used to allow and verify demand side load management control response where users are expected to change their use to respond to changes in electrical energy pricing through deferring some loads.~By knowing a home’s typical usage by device, an energy management system can perform device-specific demand response much more effectively \citep{Christensen2012}.
	
	Detail analysis of NILM data could be used for selection of pricing or incentive mechanism that maximize the effectiveness of demand response. For example, using NILM data, demand response designer can identify highest consuming appliances and their time  usages  which can be used for deriving load shift ability during peak hours.
	
	The work by Huang \etal~present an HMM-based algorithm to estimate individual household heating usage from aggregate smart meter data \cite{Huang2013}.~The authors demonstrate its application to demand response and energy audit services for thermostatically controlled heating appliances.~The work presented in \cite{He2013} demonstrate the application of real-time NILM algorithm into demand response response.
	
	Recently \cite{Batra2015a} present novel techniques that use unsupervised NILM to predict household occupancy and static household properties such as age of the home, size of the home, household income and number of occupants.
	
	\section{Energy Datasets}\label{sec:dataset}

	In the quest to design, test and benchmark a high performance energy disaggregation algorithms, NILM researchers require the availability of open-access energy consumption datasets. These dataset record the aggregate demand of the whole house as well as the ground truth demand of individual appliances and offers a real and noisy environment which can lead to more accurate algorithms design.
	
	Reference Energy Disaggregation Data Set (REDD) is the first public energy dataset released by MIT in 2011 \citep{Kolter2011}. REDD contain high and low frequency readings from 6 households in USA recorded for short period (between a few weeks and a few months). This dataset is widely used for the evaluation of NILM algorithms. 
	
	Recent years has seen the emergency of several publicly available datasets such as UK-DALE \citep{Kelly2015}, AMPDs and AMPDs2 \citep{Makonin2013,Makonin2016}, ECO dataset \citep{Beckel2014}, REFIT dataset \citep{Murray2015b} and GREED dataset \citep{Monacchi2015a}. The comparison of various publicly available dataset with their characteristics is shown in  \cref{tab:dataset}. 
	This comparison is an extension of the proposed one in \citep{Bonfigli2015} and \citep{Murray2015b} with an update of the recent published data and additional information included in some datasets.

	\section{Challenges and Future Research Directions}\label{sec:challenges}

	In the previous sections we have presented an up to date overview of NILM system and its associated methods and techniques for energy disaggregation problem highlighting the gaps and limitations.~Despite several efforts done by previous NILM researchers, there are still several challenges which need to be addressed.
	
	As a matter of fact, most of the prior NILM algorithms have been developed and tested in developed countries.~Developing countries like Tanzania offers unique characteristics such as  unreliable grid which is uncertain with blackouts and brownouts \cite{U.S.Congress1991}, different sets of appliances such as use of second-hand appliances \cite{Oyedepo2012} and different consumer behaviour.~Batra \etal~observed a significant voltage fluctuation and power outages in the data collected in India \cite{Batra2013b}.~All these factors affect the use of energy and need to be considered in the design and development of NILM methods and techniques for energy disaggregation.
	
	Second, most of the prior NILM techniques can not perform real-time disaggregation owing to algorithm complexity.~Practical NILM algorithms need to process on-line data and react in real-time to changes in the power being monitored \citep{Makonin2015}.~The few that provide real-time disaggregation utilize cloud services that introduce privacy and security concern to households data.~Future research should focus on real-time disaggregation by reducing the complexity of disaggregation algorithm.~There is also a need to explore different privacy and security techniques suitable for disaggregation algorithms  that utilize cloud services.
	
	Likewise, generalizing the learned NILM models to a new building and automatically annotating appliance events is still a problem in NILM.~Previous works rely on manually labelling appliance events after disaggregation or assumes that sub-metered ground-truth is available \cite{Iwayemi2015}.~It is very important for NILM model to be generalised to useen buildings because it is very rarely to find sub-metered data.~Future work should focus on unsupervised NILM learning algorithms that do not require human labeling of data and which can be generalized across multiple buildings.
	
	The recent study by Kelly \etal~\cite{Badayos2015} demonstrated that the use of deep learning for energy disaggregation can be generalized well to unseen buildings.~Future works should explore and investigate different unsupervised learning and deep learning algorithms for energy disaggregation.~It has also been shown that the combination of deep learning and probabilistic model such as HMM have quite promising results for energy disaggregation problem \cite{Huss2015,Mauch2016b,Zhang2016}.~Thus future works should also explore and investigate different hybrid deep-learning-HMM framework for energy disaggregation problem.
	
    In like manner several NILM algorithms have focused on computing appliance level energy breakdown and not usability or concrete application that emphasis on quantifiable energy saving in building \cite{Barker2014}.~Simply providing appliance-level energy breakdown is not a compelling application of NILM as it does not directly lead to quantifiable improvements in energy efficiency \cite{Kelly2016}.~Thus there is a need to analyze energy disaggregation data and organize it into actionable feedback that actively stimulate energy efficiency in residential building.~There is a need to expose novel NILM use-cases that use energy-disaggregation data such as how to predict electrical fires accidents or use smart-meter data for electricity theft detection just to mention a few.
    
	Equally important, energy disaggregation research have no consistent way to measure and evaluate performance and quality of NILM algorithms.~Most of the earlier works evaluate their approaches using a different set of performance metrics which make difficulties to fairly compare these algorithms \citep{Zeifman2011b}.~In addition, many of these metrics are incomparable across different algorithms for the same problem variant \citep{Barker2014} and the numerical performance calculated by such metrics cannot be compared between any two papers \citep{Batra2014a}.~Future research should also focus in standardizing NILM performance metrics.
	
	Lastly, to develop disaggregation algorithms, researchers
	require both aggregate demand per building and the ground truth demand of individual appliances data.~However, existing energy data set suffer from several problem such as incorrectly labelled sub-meters (channel labelled \textit{fridge} actually records the kitchen radio).~There is also no ground-truth labels for important NILM use-cases. For example, one important NILM use-case might be to tell people when their fridge's door seal needs replacing\footnote{http://jack-kelly.com/simulating\_disaggregated\_electricity\_data}.~Apart from that existing dataset is are from Europe, Canada, USA and India.~There are no public datasets from developing countries such as Africa.~There is thus a need to develop more data-sets from different geographical location.~However, collection of these data is very costly and time-consuming.
	
	Future research should consider developing a realistic simulators for simulating endless amount of "near-perfect", realistic  disaggregated electricity data.~Recently Chen \etal~\cite{Chen2016} present a publicly-available
	device-accurate smart home energy trace generator which generates energy usage traces for devices by combining a device
	energy model, capturing its pattern of energy usage when active and a device usage model based on its frequency,
	duration, and time of activity.~By leveraging on this simulator further research can build different statistical models for appliances in different geographical location and for 
	several usage patterns.

	\section{Conclusions}\label{sec:conclusion}
	In this work, a review of an up to date  NILM system and its associated methods and techniques for energy disaggregation problem is presented.~The review draws several conclusions; 
	
	First while several NILM techniques has been proposed for reduction and or controlling energy consumption in residential building in developed countries, there is lack of research on the use of NILM in developing countries.
	
	Second, despite the major leaps forward in the NILM field, the energy disaggregation is by no means solved.~State-of-the art algorithms still leave a lot of challenges when it comes to real-time disaggregation that is general enough to be deploy in any household.
	
	Third the standardization of NILM performance metrics, development of  efficient NILM benchmarking framework and availability of high-quality energy data set are critical for the advancement of energy disaggregation research.
	
	Lastly, despite the fact that NILM techniques promise several
	useful use-cases, its broader applications have not been fully realized.

   \section{Acknowledgments}
   The authors would like to thank Tanzania Communication Authority (TCRA) for supporting this research.

   \bibliographystyle{unsrt}
   \bibliography{bib/References_NILM}  
\end{document}